\documentstyle[12pt,amssymb]{article}

\begin{document}


\begin{center}
{\bf General Relativity in Post Independence India}\\
{Naresh Dadhich, nkd@iucaa.in}\\
{Centre for Theoretical Physics, Jamia Millia Islamia,
New Delhi 110025, India}\\
{Inter-University Centre for
Astronomy \& Astrophysics, Post Bag 4, Pune 411 007, India}

\end{center}

\begin{abstract}
The most outstanding contribution to general relativity in this era came in 1953 (published in 1955 \cite{akr}) in the form of the Raychaudhuri equation. It is in 1960s that the observations began to confront the eupherial theory and thus began exploration of GR as a legitimate physical theory in right earnest. The remarkable discoveries of cosmic microwave background radiation, quasars, rotating (Kerr) black hole and the powerful singularity theorems heralded a new canvas of relativistic astrophysics and cosmology. I would attempt to give a brief account of Indian participation in these exciting times. 
\end{abstract}




\medskip

\section{Introduction}

General relativity (GR) is certainly a theory that was born before its time as no observation, barring the little discrepancy in Mercury's orbit, or experiment demanded it. It is perhaps the unique instance where a new theory is driven entirely  by consistency of principle and concept without any reference to experiment and observation. Had 
it not been for Einstein, one would not have seriously asked for a new theory of gravitation until the discovery of quasars in early 1960s. Had the technology been not adequate or had Eddington been not so taken in by the new theory, Einstein would have had tough time to propagate his exotic theory that makes untoward demand on spacetime, which has so far remained fixed and inert, to bend -- curve. It was fortunate for Einstein and his theory that Eddington, despite observational evidence being very feeble as we now know, announced after the total solar eclipse observation expedition of 1919 that light did bend exactly as Einstein had predicted. This established GR as the most elegant and beautiful theory that prompted Paul Dirac to say that it was the greatest feat of human thought. \\

This was all very fine for Einstein to become household name over night, a worthy successor of Newton after 300 years. However where does one need this as a physical theory as it makes no contact with any observation and experiment. This was the period of about 30 years covered in Ref. \cite{jvn} -- the pre independence era in which most of the work was of mathematical nature, finding exact solutions of the Einstein equation. The period that I am to cover is when GR has turned physical and exciting. \\ 

There are many areas of active research in GR and its applications, and it is a formidable task for one to cover all of them with certain degree of competence and fairness. I would rather focus on gravitational theory related questions and some important contributions, and there are other reviews on quantum gravity \cite{abhay}, classical gravity \cite{bala}, and so on covering the entire spectrum. However, I should make a disclaimer that I won't be covering hard core relativistic astrophysics and cosmology involving CMBR, gravitational lensing, and so on. I should however mention that very important and front ranking contributions have been made in these areas.  \\

Any account of this kind is inherently fraught with the danger of earning ire and annoyance of colleagues, who consider their work significant, yet it does not find mention due to my ignorance and lack of understanding or appreciation, or sheer oversight. To them all I wish to apologize most sincerely. Jayant Narlikar on the other hand is in a very fortuitous situation because there is no one alive to question him \cite{jvn}. \\

\section{Cosmology and the Raychaudhuri Equation} 

Soon after discovery of GR in 1915, Einstein proceeded to construct a static model of the Universe and found to his surprise that it required an extension of his equation by the so called cosmological constant, $\Lambda$ (as a matter of fact had he {followed the geometric way for derivation of the gravitational equation, it would have come 
with $\Lambda$ as the constant of spacetime structure on the same footing as velocity of light \cite{lambda}). In 1922, when Friedmann obtained the nonstatic expanding Universe solution for which $\Lambda$ was no longer a necessity, it attained a very ambiguous existence until of course the discovery of accelerated expansion of the Universe in 1997 \cite{perl}. The Friedmann solution was subsequently generalized first by Lemaitre and then by Robertson and Walker, known as FLRW model of the Universe -- the standard model of cosmology. \\ 

This standard model is singular in the past where curvatures and thereby energy density diverge and this singular state was christened by Fred Hoyle as ``big-bang'' which has now become a folklore. The Universe had a singular beginning in a big-bang and since then it was expanding. FLRW model describes a very special circumstance of space (the Universe) being homogeneous and isotropic. Is the big-bang singularity a consequence of this highly symmetric circumstance or a general feature of GR equation? It is understandable that Universe may be homogeneous and isotropic now but it may not be so in early times. Here was therefore an important question whether the big-bang singularity was generic or due to symmetry? \\ 

\subsection{Raychaudhuri Equation}

There was also singularity in the Schwarzschild metric and many people including Einstein (who went on to construct  a non-singular rotating cluster in equilibrium) believed that it was genuine physical singularity.  This myth was however dispelled by Datt-Oppenheimer-Snyder solution \cite{jvn} of collapsing homogeneous dust cloud. Then came the famous Godel static rotating dust model in 1949 \cite{godel} which was singularity free. However it had undesirable features like existence of closed timelike curves, being static and large cosmological constant. Following that  Raychaudhuri thought \cite{remini} it might be possible to overcome these deficiencies but all attempts came to a naught. Ironically what he was subsequently to discover showed that such a hope could never be realized. \\ 

Then he came across a paper by Einstein and Pauli in which time-time component of Ricci tensor was written as divergence. This was the crucial hint for him and using that he could write the rate of expansion of velocity field (timelike congruence) in terms of square of shear and rotation and gravitational energy density, $\rho+3p$, without assuming any symmetry. It turned out that shear worked in line with matter to enhance gravitational attraction while rotation opposed it. This was how the famous Raychaudhuri equation was discovered. Though initially he had assumed velocity field to be geodetic and thereby missing the important divergence of acceleration term. This as well as the inclusion of null congruence was subsequently done by other authors. There is an interesting story of how did it take over two years for this remarkable discovery to be published in 1955 which should be best read in his own words \cite{remini}. In this day of Internet and online submission and referee reports, it sounds all very 
amusing. \\ 

The equation was soon recognized as governing dynamics of the Universe -- the key equation in relativistic cosmology. It would not be far from truth if I say what the Saha ionization equation is to astrophysics, the  Raychaudhuri equation is to cosmology. Further it laid the foundation for discovery of the most powerful singularity theorems in early 1960s by Penrose and Hawking stating that under reasonable physical conditions, occurrence of singularity is inevitable in GR. The equation therefore had far reaching influence and consequence. \\ 

Let me now turn to the sociological background in which the equation was discovered. He was academically lonely despite N R Sen's group in GR, mostly interested in finding new solutions. Since Raychaudhuri was taken in by the general question of cosmic singularity, he did not find much resonance with Sen, and hence he had to tread his own path. On the other hand, he was thrown out of Indian Association for Cultivation of Science for not working adequately, though he did write couple of papers, on atomic physics -- the thrust and interest area of the Director. He however found a job in Ashutosh College and then moved on to the Presidncy College.That is where he had the most remarkable and brilliant innings as a legendary teacher and mentor to galaxy of students who have done him as well as the College proud.  \\

After publication in Physical Review in 1955, it was Engelbert Schucking who gave an exposition of Raychaudhuri's paper in Jordan's relativity seminar in Hamburg. There was also an Indian student, R. L. Brahmachary attending these seminars, and it was he who wrote to Raychaudhuri that his equation was much talked about (This account is due to Jurgen Ehlers who was a fellow student of Schucking). That was the first word of appreciation that came in late 1950s. Encouraged by that and picking up courage he submitted a thesis for D.Sc. to University
of Calcutta. John Wheeler was one of the examiners and he concluded his report with the recommendation that if there was a provision for award of degree with distinction, here was the fit case. \\ 

Despite this nothing much happened except that he moved from Ashutosh to Presidency College, and it was there he played the brilliant innings as a legendary teacher and mentor to several generations of students who were well known names spread all over the world. Strangely enough, Indian scientific community remained still oblivious of the man until return of Jayant Narlikar to India in 1972. And then followed fellowships of the academies and various awards and recognitions. On this background, it stands out that in 1964, C W Misner invited him to Maryland as a Fulbright Visiting Professor.  \\

\subsection{Hoyle-Narlikar Theory} 

In mid 1960s, there was yet another very important development in terms of Hoyle-Narlikar theory of gravitation (though the place of work was Cambridge, UK but one of the authors was Jayant Narlikar who was on his return to India, like his father V V Narlikar (VVN) previously, to build a strong school of relativistic astrophysics and cosmology). Mach's principle had been an enigma for all theoretical physicists including Einstein who wanted to incorporate it in GR but could not. Firstly it means different thing to different people as there is no precise statement or formulation. One of the simplest formulation is that inertia of a particle arises out of its interaction with all other particles in the Universe. This is what Hoyle and Narlikar attempted to incorporate in their theory of gravitation which reduced to GR in the low energy limit. \\

It was an elegant and beautiful formulation but unfortunately gave rise to steady state cosmology which was not sustainable after the CMBR observation. Against all odds, Hoyle, Narlikar and Burbigde persisted by formulating quasi steady state cosmology. The Universe is cyclic and the star dust left over from the previous cycle could thermalize the star light to produce CMBR like effect. It is an ingenious and very creative attempt but looks rather far fetched on general scientific considerations. On fundamental theoretical plane, Hoyle-Narlikar theory was a brilliant attempt but it produced a cosmology which was not observationally sustainable.  \\

Besides their new theory of gravitation, in 1963 their remarkable calculations had proven the first cosmic no hair theorem \cite{hn-63} by demonstrating the stability of the de Sitter expansion against small perturbations. They also considered baryon non conservation in the particle production process \cite{jdb}. They argued that  homogeneity and isotropy of the Universe was natural for the stead state theory while it was a mystery for the big-bang theory. Had they carried on this line a little further, they could have perhaps predicted that the constant curvature density perturbation is only considtent with the steaty state cosmology.

\section{Gravitational Collapse} 

Beginning with Datt, and Oppenheimer and Snyder \cite{jvn}, it became clear that a cloud of dust under its own gravity collapsed to central singularity. In early 1960s, the mystery surrounding the Schwarzschild singularity at radius $r=2M$ (where $M$ is mass) was debunked by Kruskal when he found the coordinates that continuously traversed from outside to inside of $r=2M$ surface. That it was only a coordinate singularity arising out of a bad choice of coordinates, and not a physical singularity where curvatures diverge. However it was a surface that had very bizarre physical properties that things can fall into but nothing can come out -- defines an event horizon, no signal can come out. This was then christened as black hole by John Wheeler. \\

The question that arose was, did gravitational collapse always result into a black hole covering central singularity by event horizon -- singularity remained invisible to outside observer or could it be visible and naked to world outside? Roger Penrose proposed the cosmic censorship hypothesis that collapse always resulted in a black  hole covering the singularity. There can exist no naked singularity. Since Datt-Oppenheimer-Snyder considered homogeneous dust collapse, it was pertinent to ask what happened when dust was inhomogeneous. In early 1980s  Christodoulou was first to construct an example of naked singularity arising out of inhomogeneous collapse. This 
was a demonstration by an example that naked singularity might occur. There is no general result even today proving or disproving cosmic censorship hypothesis, however there are plethora of examples and the effort is going on. \\

Before we go further let us note one very interesting result due to Asit Banerjee. One of the important features of gravitational collapse is occurrence of shell crossing singularity. He studied this effect way back in 1967, before even the formulation of cosmic censorship hypothesis and much before flurry of activity following Christodoulou's signal of naked singularity, and demonstrated its occurrence in inhomogeneous dust collapse \cite{aban2}.  Subsequently one takes care while setting up the collapse scenario that shell crossing singularities do not occur.  \\

In all the cases of naked singularity (NS), what is envisioned is that as collapse proceeds, at some instant a singularity is formed due to inhomogeneity of distribution before formation of apparent horizon and a null ray emanating from that could come out to external observer. So far, as I know, this is not what has actually been shown.  What is shown is taking a null ray from outside and evolving back in time to end it at the singularity. It is in this latter exercise there has been a great deal of activity led by Pankaj Joshi. In particular it was shown that BH or NS result generically depending upon the initial prescription of mass function and velocity field \cite{p-d}. He has written two monographs reviewing the situation on gravitational collapse and NS \cite{pankaj}, and he is continuing on this line of work unabated. It is noteworthy that a considerable number of university people with Mathematics background have joined in this work. \\ 

In this context, one of the physical questions asked was, what was it that caused NS? What is  required for NS is that in the process of collapse there develops high concentration of matter locally in certain region while the rest of distribution being normal. This meant collapse must be incoherent which could only be caused by shear. Shear is produced by inhomogeneity. Thus shear and thereby inhomogeneity is necessary for NS, however it is not sufficient \cite{shear}. In other words incoherence caused by shear should be adequate to form NS; i.e. shear or inhomogeneity should be strong enough so that NS is formed before apparent horizon. However, it soon gets covered up as apparent horizon gulps it. \\ 

Another pertinent question is, what could be observational effects, if any of NS. It was envisaged that in the local region of NS, curvatures would be diverging and might create a fireball like phenomenon emanating high energy radiation, and soon it would be gulped by apparent horizon. This high energy radiation would therefore be very short lived like 
gamma-ray bursts (GRBs). Thus was proposed a model for GRBs, as birth cries of black holes \cite{grb}. It is a different matter that the work following this \cite{piran} use the same phrase, GRBs as birth cries of black holes, in the abstract without citing \cite{grb}. There have been some more recent work with astrophysical applications of NS and their possible observation \cite{pankaj-ramesh}. \\

It is also true that a legitimate study of NS should also take into account quantum gravity aspects as well. Of course this has to be tentative and semi-classical as there is no quantum theory of gravity. Some work in this  direction has also been done by T P Singh \cite{tp}. \\ 

Let me now come to the main question of visibility of singularity which to my mind still remains open. At singularity, curvatures are diverging, curving space very very strongly. Then how can a null ray propagate out from such a region and even if it does it must be infinitely red-shifted? That should make it practically invisible and in essence ``hidden''. The other question is, a true example of NS would, in my view, require demonstration of null ray emanating from singularity as collapse is going on yet it manages to come out. This would be a realistic situation, and I believe this still remains undiscovered. Until that happens I would not be convinced only by reverse engineering examples.  \\

A large number of people have been involved in this work, it has been comprehensively reviewed in the two books \cite{pankaj}.  \\

\section{Black Holes and Their Energetics}  

As mentioned earlier that in early 1960s Schwarzschild solution was ultimately understood as describing a static black hole and around the same time the most interesting solution was found by Roy Kerr describing a rotating black hole. The latter heralded the beginning of high energy relativistic astrophysics with its potential for illuminating physics and astrophysics around recently discovered high energy sources like quasars and AGNs \cite{bm}. It was the beginning of exciting times. \\ 

Let us however begin with the seminal work done by C V Vishveshwara in Maryland, USA for his Ph.D. thesis \cite{vishu} in which he analysed stability of Schwarzschild black hole under scalar perturbation and showed that it was indeed stable. This was a very important result for black hole physics and astrophysics. This was the pioneering work in study of quasi normal modes which subsequently occupied Chandrasekhar for a decade resulting in the monograph, The Mathematical Theory of Black Holes. It has now become a standard tool of black hole physics and giving rise to a huge amount of work and its particular applications in gravitational wave physics is very exciting. You hear ringing of black hole through quasi normal modes. \\

\subsection{Penrose Process} 

Once again Penrose came up in 1969 with an interesting proposition of extracting rotational energy from a rotating black hole -- Penrose Process (PP) of energy extraction. This is entirely driven by spacetime geometry and hence is purely geometric/gravitational process. The pertinent question is,  could this process be astrophysically applicable to high energy phenomena of quasars and AGNs? \\

The most interesting aspect of rotating black hole is that space around it also participates in rotation -- what is known as dragging of inertial frames. This means a particle in its field experiences a transverse pull in addition to usual radial pull. As particle comes closer and closer to a black hole, transverse pull becomes  irresistible which means it cannot remain fixed at a point, it has to rotate. This corresponding radius is called static surface marking limit to staticity. As particle falls below static surface then comes the limit where radial pull also becomes irresistible which marks event horizon or black hole radius. In a static black hole, there is no transverse pull hence both event horizon and static surface coincide. On the other hand, the region between them is non-vacuous and is called ergosphere, for a rotating black hole. It is this that gives rise to interesting phenomenon of Penrose Process. What happens is that in  ergosphere, a counter rotating particle can have total energy negative relative to an asymptotic 
observer. This happens because in this region spin-spin interaction energy dominates over gravitational energy, and hence if particle is counter rotating relative to black hole, its total energy will be negative. It is this property which Penrose used in envisaging the following scenario: a particle of energy $E_1$ falls from infinity into ergosphere and splits into two fragments with energies, $E_2$ and $E_3$. Let $E_2<0$ as it is allowed, and  let it fall into the hole while the other fragment comes out with enhanced energy, $E_3 = E_1 - E_2 >E_1$, because $E_2<0$. This is how energy could be extracted from a rotating black hole. Of course it is only rotational energy,  which could maximum be $29\%$, could in principle be extracted. \\

This was a very exciting prospect for powering engine of quasars. It was envisioned by John Wheeler that let a star like object graze past a supermassive black hole and split into fragments, some of which might attain negative energy and fall into the hole while their other halves come out with enhanced energy forming jets in quasars. This was indeed a fascinating possibility. A closer analysis of the process \cite {press, wald} soon showed that for a fragment to attain negative energy, relative velocity between fragments has to be $>c/2$. There is no conceivable  astrophysical process that could accelerate particles to such high speed instantaneously. PP was though a very novel process but unfortunately not astrophysically viable. Following that some variants of PP were considered but nothing of much value came. \\ 

\subsection{Magnetic Penrose Process} 

In 1985 Sanjay Wagh, Sanjeev Dhurandhar and myself argued that rotating black hole does not sit in isolation but rather sits in an environment of magnetic field. Why can't energy required for particle to attain negative energy come from electromagnetic interaction instead of kinetic energy -- the electromagnetic or magnetic version of PP. We showed that it was indeed possible and wrote a paper in The Astrophysical Journal on revival of PP for astrophysical applications \cite{wdd}. This was astrophysical rebirth of PP as magnetic Penrose Process
(MPP), so termed by Roger Blandford. We did a detailed study of this scenario and it was indeed very efficient and showed that efficiency could even exceed $100\%$ \cite{pwdd, review} as against maximum $20.7\%$ of PP. We rounded up the study by writing a review article in Physics Reports \cite{review}. \\ 

What we had shown was astrophysical viability of MPP for powering high energy sources for a discrete particle accretion which is not a scenario that obtains in reality. There was another similar competing process due to Blandford and Znajek \cite{BZ} (BZP) which came a few years earlier. The setting is the same with a rotating black hole 
sitting in magnetic field whose field lines get wound up due to inertial frame dragging and that produces a quadrapole electric field. There develops a potential difference between equatorial plane and pole which then discharges to drive energy out. This is similar to MPP and the two tend to be equivalent in strong magnetic field regime. Though MPP has geometric soul while BZP is essentially electromagnetic in nature but in strong magnetic field limit it is electromagnetic interaction that dominates hence the two tend to be the same. \\ 

For realistic study of black hole energetics, one has to study magnetohydrodynamics around rotating black hole. It was a formidable task and had to wait until adequate computational power and expertise which had come about in last few years. It is gratifying to see that MPP is the most promising powering mechanism for high energy sources like quasars and AGNs \cite{ramesh}, and its distinguishing signature is efficiency being greater than $100\%$ as predicted by us in 1986 \cite{pwdd, review}. It is a different matter that the authors of Ref. \cite{ramesh} maintain that BZP is the same as MPP. This is certainly not true because the two tend to be the same only in strong magnetic field limit, at low or zero magnetic field limit BZP would be inoperative while MPP would still work tending to PP with a reduced efficiency. At any rate more than 30 years back we were the first to consider MPP and predicted possibility of more than $100\%$ efficiency of energy extraction. It is gratifying  to see that MPP has borne out its astrophysical promise. \\

On the side let me also mention that one interesting scenario involving a white hole was also proposed as a model for high energy astrophysical sources. White hole is a time reversed black hole resembling a localized mini big-bang.  It was envisioned that white hole is like an exploding source and emerging radiation from it was shown to obey a power law spectrum consistent with the observations of the time \cite{nature}. \\

\subsection{Super-radiance} 

There is a wave analogue of PP, called super-radiance in which particle is replaced by wave. Let a scalar or  electromagnetic or gravitational wave be incident on a rotating black hole and gets scattered into transmission and reflection coefficients. The condition for super-radiance is that reflection coefficient should be greater than unity. There is an analogue of $E_2<0$ of PP for the wave frequency for the existence of super-radiance, namely wave angular momentum is counter to that of the black hole. There is a very comprehensive discussion of this phenomenon in Chandrasekhar's The Mathematical Theory of Black Hole \cite{chandra}. \\ 

One of the puzzling questions was that unlike scalar, vector and tensor waves, it turned out that Dirac wave did  not suffer super-radiance while being scattered by a charged rotating black hole. This was the question addressed by Sanjay Wagh and myself in 1985 \cite{sup-rad}, and it was shown that Dirac wave could not suffer super radiance because it violates the weak energy condition in the Kerr-Newman geometry of charged rotating black hole. That is Dirac particles in the Kerr-Newman geometry have negative energy (energy momentum dotted with timelike velocity field) and that is the cause of absence of super-radiance and also of decrease of black hole area. That is it violates the second law of black hole physics -- non-decrease of black hole area. Further it was also shown that  the breakdown of energy condition was more generally true for a rotating black hole immersed in an arbitrary electromagnetic field. This confirmed Bekenstein's conjecture that it was breakdown of the weak energy condition that was responsible for violation of Hawking's area non-decrease theorem for fermions in general. \\   

\subsection{Particle Motion Around Black Holes} 

There has been considerable work on particle motion around black holes sitting in electromagnetic field, and I could do no better than to refer to a recent comprehensive review by A R Prasanna \cite{prasanna} who is one of the leading players in this field. Though motion around a static black hole in electromagnetic field had been studied, the most interesting scenario was provided by rotating black hole in magnetic field. The property of rotation also being shared by space surrounding hole, known as inertial frame dragging effect. It is this that makes rotating black holes very exciting for it gives a new aspect to magnetic field threading the hole -- dipolar magnetic field, because of rotation generates quadrapole electric charge on the hole. All this makes study of particle trajectories, of charged as well as neutral particles, very interesting. \\

A charged particle moving in the field of rotating black hole sitting in magnetic field, would have spin-spin interaction term, $\Omega (L - qA_\phi)$, ($\Omega$ is particle angular velocity) which could be negative even when $L>0$, would we term this orbit co or counter rotating! We have to refer to locally non-rotating observer which has the frame dragging angular velocity $\omega = - g_{t\phi}/g_{\phi\phi}$. Irrespective of sign of $L$, it is motion relative to the locally non-rotating observer (LNRO) that decides co and counter rotation, it is co if particle goes ahead, and it is counter if it lags behind LNRO.  \\ 

Apart from probing geometric properties of black hole spacetimes, study of particle trajectories is also astrophysically important as it forms the basis for constructing accretion disks models around black holes. It is generally believed that high energy objects like quasars and AGNs have a black hole with an accretion disk as their powering mechanism. There is a considerable activity in this area, and a few representative names are A R Prasanna, Sandip 
Chakrabarti, Banibrata Mukhopadhyay, Tapas Das and others. \\ 

\section{Gravitational Waves} 

I would give a bit of historical perspective of how gravitational waves all began in India. It was in late 1980s when Sanjeev Dhurandhar went to Bernard Schutz at Cardiff as a postdoc and he came back fully charged to initiate work on this exciting new area. He was followed at Cardiff by B S Sathyaprakash (Sathya) and Patrick Dasgupta. That was how the former was roped in and who subsequently succeeded Schutz at Cardiff. Dhurandhar built a strong school in gravity wave data analysis and trained a large number of students and postdocs (Sathya was the first postdoc) who were spread all over the globe. On the other hand Bala Iyer at RRI focused on wave form study of gravitational wave as well as on formulation of  elaborate and comprehensive post-Newtonian schemes for analysis, and trained a number of students.  

Building up a gravitational wave detection experiment was conceived as early as late 1980s, and we had first workshop at Indore with Bernard Schutz and Alain Brillet, and experimental experts in lasers and vacuum etc from various labs participating. This was rather premature at that time, however it is gratifying to see LIGO-India is   going strong, and would soon become a reality. It is a fitting culmination of pioneering effort this dedicated group of people. \\ 

\section{Exact Solutions} 

This had been one of the favorite activities among several researchers particularly from Mathematics background. The most outstanding solution was of course Vaidya's radiating star solution which had been covered in pre independence era \cite{jvn}. Then came S Datta Majumdar solution \cite{sdm} generalizing Weyl's solution of axially symmetric charged perfect fluid distribution and he obtained a class of new solutions. He employed a coordinate transformation that made all but one equation identically satisfied, and the one was nothing but Laplace equation. This metric was also found by Papapetrou and was therefore known as Papapetrou - Majumdar metric. \\

For finding solution for fluid distribution, one has to take an equation of state or assume fall off behaviour for density or pressure, or assume metric function, a priori. Most of people resort to the latter alternative which does yield solution but its physical relevance and viability remain open question. Most often they are not  physically sustainable and that is why we have very few physically interesting exact solutions. For a comprehensive bibliography, I would refer to the two monographs \cite{exact, krasin}. Despite considerable activity among university researchers, there has not been any discovery of a transformation or ansatz as a tool for constructing new solutions. \\ 

One of the interesting prescriptions for spatial geometry is due to Vaidya and R S Tikekar in which $t=constant$ surface is given to be spheroidal instead of usual spherical, and then they constructed a fluid model for interior of a star like object \cite{vaidya-tike}. This is a physically viable star solution and it enthused good bit of work in terms of finding general solution and studying its physical properties. A fairly general solution of the Vaidya-Tikekar ansatz has been found \cite{mukh}. This is indeed a viable star interior solution.\\

Vaidya and L. K. Patel also obtained \cite{v-p} a solution for radially out flowing radiation in the Kerr geometry for describing a radiating Kerr black hole. Despite the powerful singularity theorems, in 1990 Senovilla discovered \cite{seno} a singularity free cylindrical cosmological solution. How did it defy the theorems? Though all the assumptions leading to the theorems looked physically reasonable except the one of existence of trapped surface, and that is the one Senovilla's solution violated. Following that a number of singularity free solutions were obtained with the stiff fluid equation of state $\rho=p$ \cite{dpt} as well as a spherical cosmological model was constructed \cite{d-r}. \\   

Then there was a general result due to S R Maiti \cite{maiti} who showed without solving Einstein equation that a spacetime in absence of shear and rotation was always spherically, plane or hyperbolically symmetric. He used only equation of motion and evolution equation for kinematic tensors, and had however tacitly assumed without mentioning Weyl curvature to be non-zero. \\ 

It is generally believed that the Brans-Dicke theory reduces to GR for large $\omega$ limit, Narayan Banerjee and Somasri Sen showed that this is only true for tracefull matter distribution \cite{nar-som}. That is for trace free matter like radiation, infinite $\omega$ limit Brans-Dicke theory does not go to GR. \\  

The frequency shift for a light emitted by a collapsing object would naturally have both Doppler and gravitational contributions, that was what shown by Asit Banerjee \cite{aban1}. \\ 

S Banerji found an interesting method of operational determination of curvature coordinates for the Schwarzschild spacetime \cite{sban} which was previewed for its novelty by the Nature Editor. \\

\section{Beyond GR}

There are two most pressing motivations for seeking beyond GR. One, GR is singular at both ends, end
state of a massive star and beginning or end of the Universe. Second, like all other forces, gravity
should also accord to quantum formulation simply because matter that produces it is deep down quantum
in character, and so should be spacetime which mediates gravitational dynamics. Naively spacetime bends or
curves like a material object, it ought to have discrete micro structure deep down so that it can bend. Thus
at the fundamental level, it is therefore pertinent to look beyond GR for a new understanding or resolution of
these problems. The measure of the question is in the fact that it is not for nothing that the best of minds
have been breaking their heads for over half century, yet the problem is so formidable that we are nowhere near
solution. \\

Einstein formulated a guiding principle in the form of Equivalence Principle for seeking beyond Newton, and
following that he arrived at the most revolutionary theory -- GR. I believe we seriously need an analogue of
Equivalence Principle to seek beyond Einstein. Unfortunately we have not yet been able to formulate one. If we
reflect on what is that we should take from Einstein's way of doing science? The clear message that seems to emerge is,
seek out synthesis of all universal concepts and phenomena. He came to special relativity by seeking synthesis of the
two most primary universal entities space and time, and then came to GR by synthesizing the universal force of gravity with
spacetime. Is there anything else which is also universal, and hence has to be brought into spacetime fold? The answer is
quantum principle, which is universal as it applies to everything, and so it must be synthesized with spacetime. That is what is
required for completion of quantum theory and that would be a quantum theory of spacetime. Since gravity is already in the
spacetime fold, it would also be a theory of quantum gravity. This shows how involved and all encompassing is this question. \\  

\subsection{Gravity and Thermodynamics} 

In 1992, in International GR Society meeting in Cordoba, Ted Jacobson \cite{Jacobson:1995ab} came out with the brilliant observation that Einstein vacuum equation near a null horizon takes the form of the First Law of thermodynamics. It is strongly indicative of a connection between gravity and thermodynamics. This augurs wonderfully well with the Einsteinian message that all universal things should be related. Like gravity, thermodynamics is also universal and hence they should be related. This gave rise to the view that gravity may not be a fundamental force but it is, like thermodynamics, a bulk property. Gravity is not a fundamental force but an emergent force. Like some others, T Padmanabhan is strongly taken in by this view and has built a huge body of work with his students and collaborators \cite{Padmanabhan:2002sha,Paranjape:2006ca,Padmanabhan:2009vy,Padmanabhan:2013nxa,Chakraborty:2014rga,Chakraborty:2015wma,
Chakraborty:2015aja,Chakraborty:2015hna}. This is indeed a very impressive and significant contribution to the most pressing and challenging problem of the day. \\

We all know, rather relativists pride in the fact, that gravity is not like any other force, it is different as no other force makes such a demand on spacetime to curve to accommodate its dynamics. Gravitational field is sitting in space curvature and that is why we cannot capture it with the the usual tools we have in armory for handing other forces. Space curvature is like thermodynamics, though defined locally but effective at large - essentially a bulk property, and it is perhaps that
makes gravity to take thermodynamical form. That however does not make it entirely an emergent force, what it seems to indicate is that it also has a thermodynamical like manifestation. It is like photon which is both particle as well as wave. Similarly gravity is also perhaps both, fundamental as well as emergent. Like photon it is self dual (I am using self dual phrase in non-technical sense). \\

\subsection{Brane World Gravity} 

There was another interesting probing in terms of the brane world gravity. For relativists, the most pleasing aspect of it was that it was curvature of bulk space that made extra dimension effectively compact. This was a welcome departure from string theorists prescription of compact extra dimension which was not demanded by any gravitational property. The
main motivation for this attempt was however to understand mass hierarchy in particle physics which however did not go far enough. However it provided an interesting way to explore possible effects of higher dimensional gravity on four dimensional world. About a decade and half back brane world gravity was formulated by L. Randall and R. Sundrum \cite{RS} and it instantly caught fire resulting in huge number of papers at an amazing speed. The model envisages on the lines of string theory that all matter fields remain confined to four dimensions, called 3-brane or brane while gravity can propagate in extra dimension, called bulk. The usual four dimensional spacetime is a hypersurface  in higher dimensional bulk spacetime. Gravitational equations in the two are related by the junction conditions driven by the Gauss-Codazzi equation relating curvatures in the two spaces. It turns out that the equation on the brane is not closed as it has projection of bulk Weyl curvature on the brane. It is therefore formidable to 
solve bulk and brane equations simultaneously. In here extra dimension is effectively made compact by curvature of bulk and it could be shown that propagation of gravity in bulk is not massless free but rather massive like Kaluza-Klein  and hence it cannot penetrate deep enough. In particular bulk is taken as anti-de Sitter and that produces an $1/r^3$ correction to Newtonian potential on the brane. Similarly Schwarzschild-AdS black hole bulk {\bf harbors} FLRW cosmology on the brane. \\

It was shown that Reissner-Nordstrom metric of charged black hole was solution of brane vacuum equation and it  could therefore describe a black hole on the brane \cite{brane-bh}. This was the first brane black hole solution which may not be entirely satisfactory, yet I believe it captures the essential aspects. Here charge in the metric is not Maxwell electric charge but gravitational charge due to projection of bulk Weyl curvature on the brane. It could be termed Weyl charge and it appears in the metric with opposite sign ($-W^2/r^2$ instead of $+Q^2/r^2$).  In the early years of this century, there was intense activity in the brane world gravitational model with its applications to cosmology and astrophysics in various modes and situations. There was a fair degree of Indian  participation in this venture by Sayan Kar and others. \\

\subsection{Bulk and Surface Terms in Action} 

As we said earlier that gravity is different from other fields and a telling demonstration of this fact is in gravitational action itself which is second order in derivative and strangely yet on variation it yields second order equation. What did we then vary? What happens is that variation of the entity involving second derivative, Ricci tensor, is cast as a divergence and hence converted into a surface term and thrown out. It is variation of the remainder; i.e. $R_{ab}\delta(\sqrt{-g} g^{ab})$ that gives the Einstein equation. Surface terms are though trivial for classical physics, they turn non-trivial for quantum considerations. There is a good body of work on manipulation and significance of surface terms. \\ 

T. Padmanabhan had first shown that a clever manipulation of surface term also yields the same gravitational equation \cite{Mukhopadhyay:2006vu}. That is bulk and surface are kind of dual to each other, and there must therefore exist a relation between the two. That is what he has established. This is indeed very important result which I believe  encodes holography -- a relation between bulk and surface. I would though like to hazard a speculation that  holography is perhaps an inherent gravitational property and not a new principle. \\ 

There are three kinds of surface terms known as Euler, Pontryagin and Nieh-Yan topological invariants. Euler is the familiar one which appears in variation of Einstein-Hilbert action. Ghanashyam Date, Romesh Kaul and Sandipan Sengupta have analysed the role of the Nieh-Yan invariant in canonical gravity and provided a topological interpretation of Barbero-Immirzi parameter which also appears in loop quantum gravity \cite{kaul etal}. Subsequently, Kaul and Sengupta have also studied how the Hamiltonian theory of gravity is modified when all three of these topological densities are included in the four dimensional gravity Lagrangian. In a more recent work \cite{kaul-sen} based on torsional instantons, these two authors have conclusively demonstrated that the Barbero-Immirzi parameter in fact affects the vacuum structure of quantum gravity in exactly the same way as does the theta angle in gauge theories. All this, though very involved for any further discussion here, appears quite interesting and insightful and holds promise for new direction. \\

\section{What Made The Difference}

I would like to recount some of the initiatives, institutions and organizations that made difference in growth and strengthening of GR research in the country.

\subsection{Summer School}

As mentioned earlier, this was also the beginning of exciting times in GR and cosmology with black holes, big-bang  and CMBR becoming the talk of the town. To work in these areas, one has to have a good grounding in both physics and mathematics. Since GR was not part of physics teaching programme in most universities, students remained untouched of exciting GR research. The best way to remedy the situation was to start a summer school series for doctoral and master students for conveying this excitement and motivating them for participation. At the instance of P C Vaidya, A K Raychaudhuri, J V Narlikar, C V Vishveshwara and A R Prasanna, and with the blessings of D S Kothari, UGC agreed to support the summer school programme. It soon had an effect in more physically oriented GR work being published in reputed journals. \\ 

\subsection{IAGRG}

In 1969, P C Vaidya organised the first meeting in GR at Ahmedabad at the occasion of V V Narlikar turning 60 years of age, and an association was formed, Indian Association of General Relativity and Gravity (IAGRG) with VVN as its first President. Fred Hoyle had attended the foundation meeting and he was made the first Honorary Fellow of the Association. The next meeting of IAGRG as a body was held in 1971 at Banaras Hindu University, Varanasi and since then it had been holding its meetings with an interval of year and half, firstly as a part of big meetings like Indian Mathematical Society. From 1980s it started meeting of its own with similar or a bit increased frequency. The relativity community got organised as a professional academic body.  \\ 

\subsection{ICGC}

One of the other handicaps young researchers faced was lack of interaction with leading experts and exposure to world outside as there was not enough support available for attending international conferences abroad. It was thought why not do the reverse, bring the experts from abroad here by means of an international conference. This was how International Conference on Gravitation and Cosmology (ICGC) series was conceived. At the initiative of Jayant Narlikar, the first meeting was held in 1987 hosted by TIFR at Goa with a good world leaders contingent consisting of Fred Hoyle, Roger Penrose, Ted Newman, Karel Kuchar, Juergen Ehlers, Bryce DeWitt, James Hartle, Robert Wald, Abhay Ashtekar and others. Since then this conference is being held at four yearly interval at various places in the country. The last one in 2011 was again organised by TIFR at Goa, marking 25 years of the Series, with large number of distinguished colleagues from all over the world. The next meeting in December 2015 is being hosted by IISER, Mohali. This series has got established as one of the regular international meetings. \\ 

Now the times have changed, there are many more meetings, at times too many for comfort as well as travel to international meetings and otherwise for exchange and collaboration has become common place. \\ 

In 2001, another small compact group meeting series, Field Theoretic Aspects of Gravity (FTAG), was started with its first meeting held at IUCAA. The idea was to have an intense discussion, not in the seminar talk but rather in discussion mode, among the compact group working on field theoretic aspects on new work and direction being explored. This series is also going on with a frequency roughly of an year and half with support from IUCAA, IMSc and other institutes. \\  

\subsection{IUCAA} 

With the initiative of Yash Pal, UGC Chair  and J V Narlikar, IUCAA was established in December 1988 for  strengthening and growth of astrophysics and astronomy teaching and research in the universities. The challenge of creating a world class institute with this new mandate was taken up by the latter. IUCAA has just celebrated 25 years of its meaningful journey and has made a mark as one of the leading academic centers in the world. It  supports university teachers and students in their research through its Associateship programme by providing hassle free excellent academic facilities for their research. This has resulted in enhanced research output both in quantity and quality from the university sector. In all IUCAA's programmes, university participation is an essential part. Many IUCAA Associates are relativists and their interaction with IUCAA has helped them in broadening their research base as well as in taking up physically interesting problems.  \\

\subsection{Research Schools} 

1970s saw nucleation of activity in relativistic astrophysics and cosmology in various institutes, at TIFR around J V Narlikar and S M Chitre, RRI around C V Vishveshwara and B R Iyer, PRL around A R Prasanna, and in some IITs and universities. The group at TIFR grew into one of the strongest schools in astrophysics and cosmology with formidable impact and great international recognition. This was a wonderful carrying forward of the legacy of the Varanasi and Kolkata schools of the beginning years. \\ 

\section{Conclusion} 

In the post independence era, research in gravitational physics has seen a sea change in the country. In almost all areas, there is a perceptible activity while in some others like cosmology and high energy astrophysics it is at the frontiers. Of course the centre of action is in application of GR as a theory of gravitation to various astrophysical and cosmological problems. This is mark of the fact that the theory is now fully matured as a theory of gravitation and is well grounded in solid physical base. There are however some formal open questions which are quite involved and complex but not so bothersome for practical use. \\ 

In recent times, the focus is undoubtedly on experimental activity in detection of gravitational wave, precision tests of the inverse square law and probing GR in high energy regime through astrophysical and cosmological observations in various space and ground based missions. LIGO-India would herald a new era of gravitational physics in the country as it would make us an active participant in this exciting global effort which has the potential of opening a new `eye' to the Universe. It would be wonderful to have a share in creation of that eye. So very often we miss the bus, I hope this time we don't, and this is the right time to get onto it. \\ 

Let me end it with a reflection. Had Raychaudhuri had a good mathematical environment around him, could he or someone around him not discovered the powerful singularity theorems? The same could be said of Magnetic Penrose Process and quasi normal modes of black holes. The answer is that for a full bodied growth of an idea, it is crucial to have conducive land in terms of intellectual and technical competence, as well as required technological equipment. Apart from the lack of adequate environment, both in man and machine, we seem to be happy in establishing a thing in principle but we desperately lack perseverance to take it to its logical and fruitful conclusion. It is the mixture of both of these trends that has been responsible for our missing the bus. I hope that we as a scientific community have matured and committed enough that we would overcome the perennial habit of missing the bus. \\

P C Vaidya and A K Raychaudhuri were the living examples of people who pursued scholarly work under most adverse and challenging circumstances -- perhaps they couldn't have done anything else that could have fulfilled them. It is not for nothing that we revere them so fondly with great warmth and intensity. We are all riding on their shoulders with utmost gratitude and indebtedness. It is to that noble and prestine spirit of truth seeking I dedicate this story. \\

\section*{Acknowledgement}

I thank Narayan Banerjee, Sandipan Sengupta and Sumanta Chakraborty for providing the releveant references. I also thank Jayant Narlikar for advice and John Barrow for pointing out the first cosmic no hair theorem due to Hoyle and Narlikar. \\


\begin{thebibliography}{90}
\bibitem{akr} 
A. K. Raychaudhuri, Phys.Rev. {\bf 98}, 1123 (1955)

\bibitem{jvn} 
J. V. Narlikar, Current Science {\bf 109},    (2015)

\bibitem{abhay} 

A. Ashtekar, Current Science {\bf 109},   (2015)

\bibitem{bala} 
B. Iyer, Current Science {\bf 109},   (2015)

\bibitem{lambda} 

N. Dadhich,  Int. J. Mod.Phys. {\bf D20}, 2739 (2011) 

\bibitem{perl} 

S. Perlmutter et al, Astrophys. J. {\bf 483}, 566 (1997); Nature {\bf 391}, 51 (1998).

\bibitem{godel} 

K. Godel, Rev. Mod. Phys. {\bf 21}, 447 (1949).

\bibitem{remini}

A. K. Raychaudhuri, in Raychaudhuri Equation at cross-roads (eds Naresh Dadhich, Pankaj Joshi and Probir Roy), Pramana {\bf 60}, 3 (2007)

\bibitem{hn-63}

F. Hoyle, J. V. Narlikar, Proc. Roy. Soc. {\bf A263}, 1 (1963) 

\bibitem{jdb}

F. Hoyle, J. V. Narlikar, Proc. Roy. Soc. {\bf A290}, 143 (1966) 

\bibitem{p-d}

P S Joshi, I H Dwivedi, Phys. Rev. {\bf D47}, 5357 (1993)

\bibitem{aban2} 

A. Banerjee, Phys. Rev. {\bf 150}, 1086 (1996)

\bibitem{pankaj} 

P S Joshi, Gravitational collapse and spacetime singularities, Cambridge University Press, (2008) 

\bibitem{shear}

P S Joshi, R Goswami, N Dadhich, The critical role of shear in gravitational collapse, arxiv:gr-qc/0308012

\bibitem{grb}

P S Joshi, N Dadhich, R Maartens, Mod. Phys. Lett. {\bf A15}, 991 (2000)

\bibitem{piran}

T Piran, Gamma-ray Bursts -- when theory meets observations, arxiv:astro-ph/010134

\bibitem{pankaj-ramesh}

P S Joshi, D Malaferina, Ramesh Narayan, Class. Quant. Grav. {\bf 31}, 015002 (2014) 

\bibitem{tp}

R Tibrewala, S Gutti, T P Singh, C Vaz, Phys. Rev. {\bf D77}, 064012 (2008) 

\bibitem{bm} 

B. Mukhopadhyay, Current Science {\bf 109}, .. (2015)

\bibitem{vishu}

C V Vishveshwara, Phys. Rev. {\bf D1}, 2870 (1970); Nature {\bf 227}, 936 (1970) 

\bibitem{press} 

J M Bardeen, W Press, S Teukolsky, Ap. J. {\bf 178}, 347 (1972) 

\bibitem{wald}

R M Wald, Ap. J. {\bf 191}, 231 (1974) 

\bibitem{wdd}

 S M Wagh, S V Dhurandhar, N Dadhich, Ap. J. {\bf 290}, 12 (1985); Erratum, Ap. J. {\bf 301}, 1018 (1986)

\bibitem{pwdd} 

S Parthasarthy, S M Wagh, S V Dhurandhar, N Dadhich, Ap. J. {\bf 307}, 38 (1986) 

\bibitem{review} 

S.M. Wagh and N. Dadhich, Phys. Rept. {\bf 183}, 137 (1989). 

\bibitem{BZ} 

R D Blandford, R L Znajek, Mon. Not. Roy. Astro. Soc. {\bf 179} 433 (1977)

\bibitem{ramesh}

R Narayan, Energy extraction from spinning black holes:Relativistic jets, in Relativity and Gravitation 100 years after Einstein in Prague, eds, J Bicak, T Ledvinka (Springer, 2013) 

\bibitem{nature}

J V Narlikar, K Apparao, N Dadhich, Nature {\bf 251}, 591 (1974) 

\bibitem{chandra}

S Chandrasekhar, The Mathematical Theory of Black Hole (Oxford University Press, 1983)

\bibitem{sup-rad}

S M Wagh, N Dadhich, Phys. Rev. {\bf D32}, 1863 (1985) 

\bibitem{prasanna} 

A.R. Prasanna, J. Electromagn. Waves Appl. {\bf 29}, 283 (2015). 

\bibitem{sdm}

S Datta Majumdar, Phys. Rev. {\bf 72}, 390 (1947) 

\bibitem{exact}

H Stephani, D Kramer, M Maccallum, C Hoenselaers, E Herlt, Exact solutions of Einsteins' field equations (Cambridge,
 University Press, 2013)

\bibitem{krasin}

A Krasinski, Inhomogeneous cosmological models, (Cambridge,  University Press, 1997)

\bibitem{vaidya-tike}

P C Vaidya, R Tikekar, J. Astrophys. Astrono. {\bf 3}, 325 (1982)

\bibitem{mukh}

S Mukherjee, B C Paul, N Dadhich, Class. Quant. Grav. {\bf 14}, 3475 (1997)

\bibitem{v-p}

P C Vaidya, L K Patel, Phys. Rev. {\bf D35}, 1491 (1973)

\bibitem{seno}

J M M Senovilla, Phys. Rev. Lett. {\bf 64}, 2219 (1990)

\bibitem{dpt}

N Dadhich, L K Patel, R Tikekar, Pramana {\bf 44}, 303 (1995)

\bibitem{d-r}

N Dadhich, A K Raychaudhuri, Mod Phys. Lett. {\bf A14 }, 2135 (1999)

\bibitem{maiti}

S R Maiti, Gen. Relativ. Grav. {\bf 16}, 297 (1984)

\bibitem{nar-som}

N Banerjee, S Sen, Phys. Rev. {\bf 56}, 1334 (1997) 

\bibitem{aban1}

A Banerjee, Proc. Roy. Soc. {\bf 91}, 784 (1967) 

\bibitem{sban}

S Banerji, Nature, Phys. Sc. {\bf 239}, 140 (1972) 

\bibitem{Jacobson:1995ab}

T. Jacobson, Phys. Rev. Lett. {\bf 75}, 1260 (1995) arXiv:gr-qc/9504004.

\bibitem{Padmanabhan:2002sha}

T.~Padmanabhan, Class. Quant. Grav. {\bf 19}, 5387 (2002) arXiv:gr-qc/0204019.

\bibitem{Paranjape:2006ca}

A. Paranjape, S. Sarkar, and T. Padmanabhan, Phys. Rev. D {\bf 74}, 104015 (2006) arXiv:hep-th/0607240. 

\bibitem{Padmanabhan:2009vy}

T. Padmanabhan, Rept. Prog. Phys. {\bf 73}, 046901 (2010) arXiv:0911.5004.

\bibitem{Padmanabhan:2013nxa}

T. Padmanabhan, Gen. Rel. Grav. {\bf 46}, 1673 (2014) arXiv:1312.3253.

\bibitem{Chakraborty:2014rga}

S. Chakraborty and T. Padmanabhan, Phys. Rev. D {\bf 90}, 124017 (2014) arXiv:1408.4679.

\bibitem{Chakraborty:2015wma}

S. Chakraborty, JHEP {\bf 08}, 029(2015) arXiv:1505.07272.

\bibitem{Chakraborty:2015aja}

S. Chakraborty, K. Parattu, and T. Padmanabhan,  JHEP {\bf 10}, 097 (2015) arXiv:1505.0529.

\bibitem{Chakraborty:2015hna}

S. Chakraborty and T. Padmanabhan, arXiv:1508.04060.

\bibitem{RS} L. Randall and R. Sundrum, Phys. Rev. Lett. {\bf 83}, 3370 (1999) arXiv:hep-ph/9905221.

\bibitem{brane-bh} 

N. Dadhich, R. Maartens, P. Papadopoulos, and V. Rezania, Phys. Lett. {\bf B487}, 1 (2000) arXiv:hep-th/0003061.

\bibitem{Mukhopadhyay:2006vu}

A. Mukhopadhyay and T. Padmanabhan, Phys. Rev. D {\bf 74}, 124023 (2006) arXiv:hep-th/0608120.
  
\bibitem{kaul etal} 

G. Date, R.K. Kaul and S. Sengupta, Phys. Rev. D {\bf 79}, 044008 (2009) arXiv:0811.4496.
\bibitem{kaul-sen} 

R. K. Kaul and S. Sengupta, Phys. Rev. {\bf D85} (2012) 024026; Phys. Rev. {\bf D90} (2014) 124081




\end{thebibliography}
\end{document}